# Large Bi-2212 single crystal growth by the floating-zone technique

J. S. Wen*, Z. J. Xu, G. Y. Xu, M. Hücker, J. M. Tranquada and G. D. Gu

*Condensed Matter Physics and Materials Science Department, Brookhaven National Laboratory, Upton, New York, 11973, USA*

**Abstract**
Effects of the growth velocity on the crystal growth behavior of $Bi_2Sr_2Ca_1Cu_2O_x$ (Bi-2212) have been studied by the floating-zone technique. The results show that a necessary condition for obtaining large single crystals along the c-axis is that the solid-liquid interface of a growing rod maintains a stable planar growth front. The planar liquid-solid growth interface tends to break down into a cellular interface when the growth velocity is higher than 0.25 mm/h. Single crystals of up to $50\times7.2\times7$ mm$^3$ along the a-, b- and c-axes respectively have been cut from a 7.2 mm diameter rod obtained with optimum growth conditions. $T_c^{onset}$ is 91 K as determined from magnetization measurements on as-grown crystals. Optical polarization microscopy and neutron diffraction show that the quality of the large single crystals is good.

**Keywords:** A1. Crystal growth, A2. Floating zone technique, B1. Cuprates, B2. Superconducting materials

It is not a great challenge for crystal growers to obtain thin, plate-like crystals of $Bi_2Sr_2Ca_1Cu_2O_x$ (Bi-2212), with a thickness of up to 0.1 mm along the c-axis, by the flux method and the Bridgman method [1-6]. One can also use the traveling-solvent floating-zone (TSFZ) technique to grow Bi-2212 single crystals [7-10]. The latter work has shown that Bi-2212 single crystals with a thickness of less than 0.2 mm along the c-axis are easily obtained under various growth conditions and initial compositions. Gu *et al.* have succeeded in using the TSFZ technique to grow relatively large Bi-2212 single crystals and showed that the optimal starting composition was $Bi_{2.1}Sr_{1.9}Ca_{1.0}Cu_{2.0}O_x$ [11,12]. They reported that while deviating from the optimal starting composition, the planar interface broke into cellular interface. However, it is still very difficult to grow Bi-2212 single crystals with size large enough for neutron scattering measurements. In this paper, we report the effect of the growth velocity on Bi-2212 crystal growth behaviors. The optimum growing conditions for preparing large single crystals with a thickness of up to 7 mm along the c-axis are presented.

An infrared radiation furnace equipped with two ellipsoidal mirrors and two 500 W halogen lamps placed in the focus of the mirrors was used for the crystal growth. Powders of $Bi_2O_3$, $SrCO_3$, $CaCO_3$ and CuO (99.99%) in their metal ratios were mixed, ground in an agate mortar and calcined for 48 h at 810°C. The calcined powders were then reground and calcined a second time. The reground powders were placed in a rubber tube and hydrostatically pressed under 6000 kg/cm$^2$. The pressed rods were sintered for

---

* Corresponding author: jwen@bnl.gov, 631-344-3714.

72 h at 860°C for use as feed rods and at 840°C for solvent materials. The sintered rods were then pre-melted at a velocity of 25 mm/h; this is necessary for obtaining a stable molten zone during the crystal-growth cycle. The experimental conditions are summarized in Table 1. All experiments were performed in air, and the temperature gradient at the solid-liquid interface of the seed rods was not measured. In order to reveal the morphology of the growth front, the molten zone between the feed rod and the seed rod was quenched.

Table 1 Growth conditions for Bi-2212 crystals

| | | | |
|---|---|---|---|
| Feed rod diameter | 8 mm | Seed rod diameter | 7.2 mm |
| Feed rod rotation | 30 rpm | Seed rod rotation | 30 rpm |
| Feed rod velocity | 0.25 mm/h | Seed rod velocity | 0.25 mm/h |
| Feed rod composition | $Bi_{2.1}Sr_{1.9}Ca_{1.0}Cu_{2.0}O_x$ | Solvent composition | $Bi_{2.5}Sr_{1.9}Ca_{1.0}Cu_{2.6}O_x$ |

The microstructure of the polished sections was examined by an optical polarization microscope under which differently oriented crystals appear with different colors. Single crystals were identified and extracted based on the images taken with the microscope. The quality of the single crystals was analyzed by neutron scattering measurements and the static magnetic susceptibility of the as-grown crystal was measured with the Magnetic Properties Measurement System (MPMS).

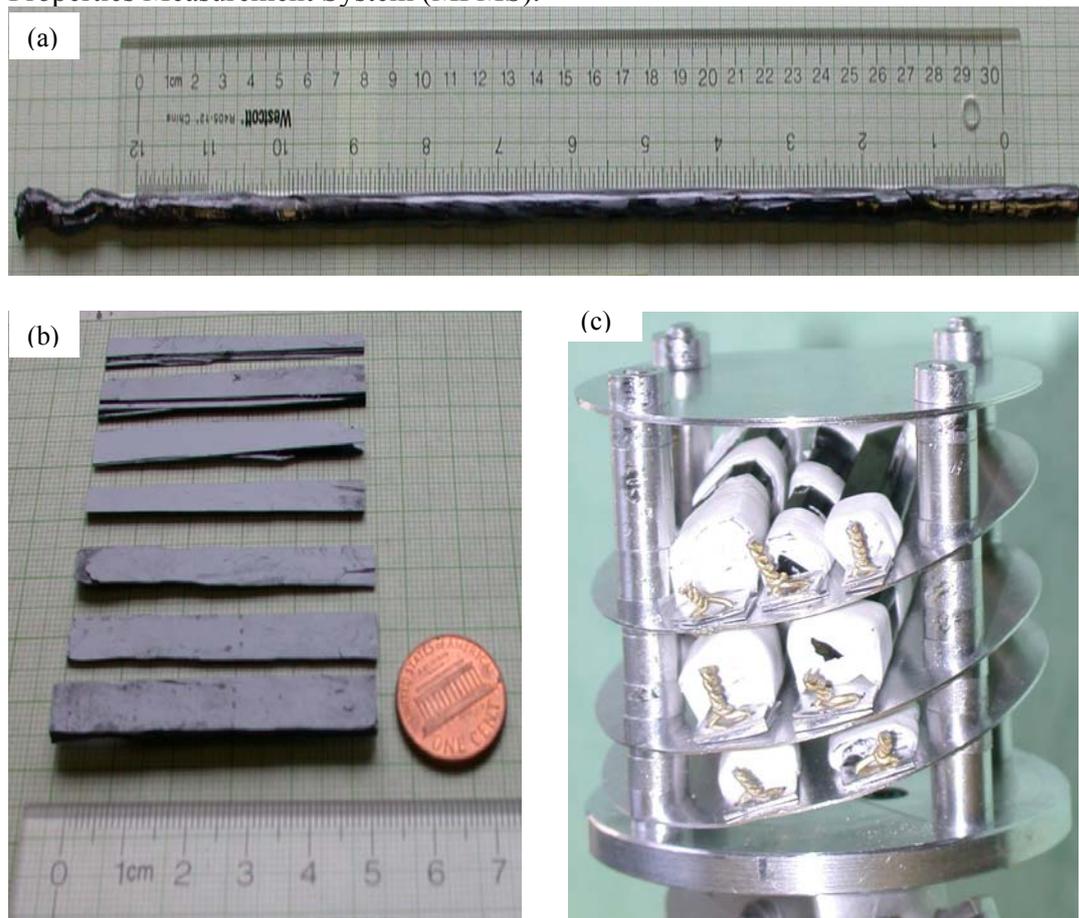

Fig. 1, (a) As grown crystal rod, 32 cm long. (b) Single crystals cut from the as grown rods. (c) Co-aligned single crystals for neutron scattering measurements.

Fig. 1 shows the crystals grown with the starting composition $Bi_{2.1}Sr_{1.9}Ca_{1.0}Cu_{2.0}O_x$. Fig.1(a) is an entire as-grown crystal rod, with a-axis along the growth direction. For the initial part of the rod (0 cm- 6 cm, at left in Fig. 1(a)), the growing conditions are not optimized, and the growth is not stable. Therefore, the size of the crystals is not large and the composition may not be right (being affected by the solvent composition). At a certain point, the growth becomes stable and the size of individual crystals optimizes. Fig.1(b) shows single crystals cut from the as grown rods. They are 50 mm along a-axis and 7.2 mm along b-axis. The largest crystal has a thickness of 7 mm along c-axis. For the neutron scattering experiments we have co-aligned the crystals using the Laue X-ray back reflection technique, and the resulting ensemble of crystals are shown in Fig. 1(c).

Polarization photographs of the longitudinal and cross sections of the as-grown rods at the velocity of 0.5 mm/h and 0.25 mm/h are shown in Fig. 2.

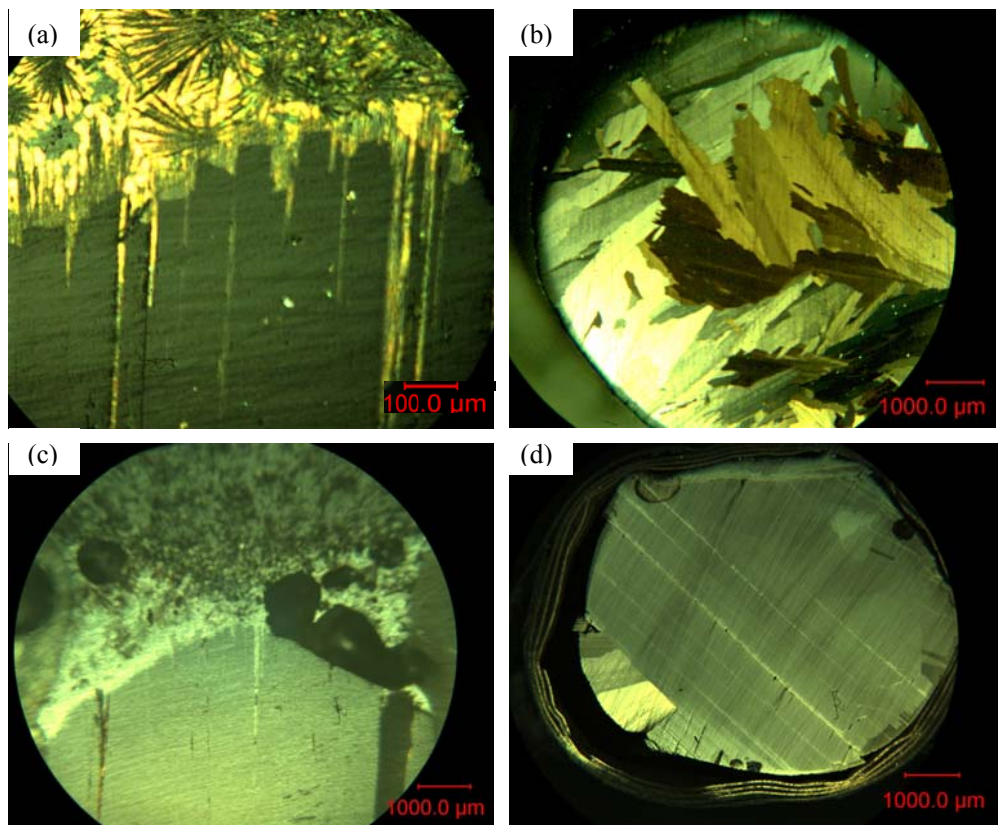

Fig. 2, Longitudinal and cross section of crystal grown with a velocity=0.5 mm/h, (a) and (b); velocity=0.25 mm/h, (c) and (d).

With a high growth velocity such as 0.5 mm/h, the solid-liquid interface is cellular as shown in Fig. 2(a) which is the longitudinal section of a quenched rod. The cross section of the as-grown rod, Fig. 2(b), exhibits many small crystals. When using a lower velocity of 0.25 mm/h, the planar interface can be maintained, as shown in Fig. 2(c), and large single crystals can be obtained (Fig. 2(d)). The growth front morphology is very sensitive to the growth velocity, which is critical for obtaining large single crystals along the c-axis. Apparently, a low velocity is very useful in maintaining a smooth growth front of a rod.

The magnetization was measured under zero field cooling (ZFC) with a field of 2 Oe along the c-axis and the result is shown in Fig. 3. It shows that the sample is homogeneous with $\Delta T_c(10\%-90\%)=2K$. Based on the $T_c$ of 91K, we believe that the sample is optimally doped. Neutron diffraction measurements, performed at the NIST Center for Neutron Research, confirm the quality of the individual crystals. It is important to note that neutrons probe the entire crystal volume. Fig. 4 shows a typical rocking curve for the (0,0,10) Bragg peak. The full-width-at-half-maximum (FWHM) of ~1° shows that the mosaic distribution is approximately Gaussian and reasonably narrow.

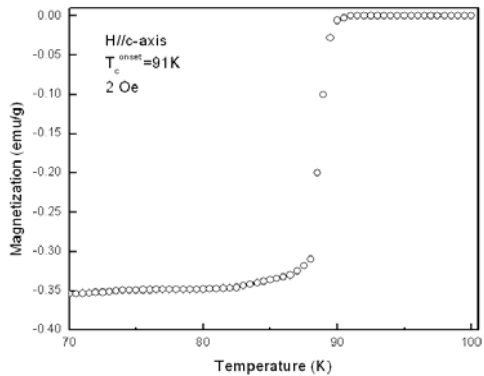 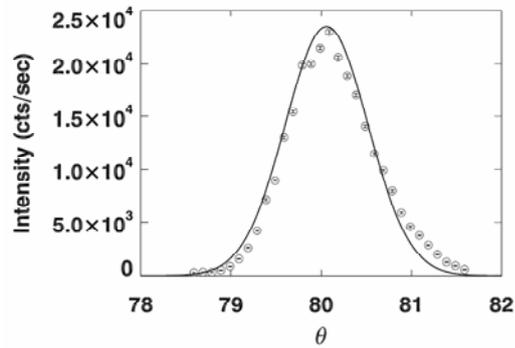

Fig. 3, Magnetization measurement of the single crystal.

Fig. 4, Rocking curve for the (0,0,10) Bragg peak.

One problem we encountered is that often there are non-continuous parallel micro-cracks along *b* direction within a single crystal, as can be seen in Fig. 2(d). The micro-cracks are useful to study the grain boundary, for we can determine the grain boundary angle according to the micro-cracks between two crystals. However, the non-continuous micro-cracks are an obstacle for transport measurements along the c-axis. The larger the c-axis thickness of a crystal, the more serious is the micro-crack problem. The intrinsic cause for the micro-cracks in crystals is that only a very weak linking force exists between the two Bi-O layers. The external cause for the micro-cracks is that there is a large thermal stress due to high radial and axial temperature gradients during the crystal growth, and also there is mechanical stress produced by cutting crystals from the as-grown rod.

In conclusion, Bi-2212 single crystals with size as large as $50\times7.2\times7$ mm$^3$ have been grown by the TSFZ technique with a velocity of 0.25 mm/h and a starting composition of $Bi_{2.1}Sr_{1.9}Ca_{1.0}Cu_{2.0}O_x$ in the feed rod. For the as-grown crystals, $T_c^{onset}$ = 91 K and $\Delta T_c(10\%-90\%)=2K$, which shows that the single crystal is homogeneous. Optical polarization microscopy and neutron measurement show that the crystal is single and of high quality. For growing large single crystals along c-axis, it is necessary that the smooth solid-liquid interface of an as-grown rod be maintained. If the smooth growth interface breaks down into a cellular interface, the thickness decreases rapidly. A starting composition of $Bi_{2.1}Sr_{1.9}Ca_{1.0}Cu_{2.0}O_x$ and a low velocity as 0.25 mm/h are helpful for preventing the smooth interface from breaking down into the cellular interface.

**Acknowledgements**

The work at Brookhaven National Laboratory is supported by the U.S. Department of Energy's Office of Science under Contract No. DE-AC02-98CH10886. We also thank the beamline scientists Ying Chen and Jeff Lynn at BT-7, NCNR, NIST for help during the neutron measurements.